\documentclass[onecolumn,preprintnumbers]{revtex4}
\usepackage{amsmath,amssymb,graphics,epsfig,subfigure}
\usepackage{color}
\usepackage[colorlinks,linkcolor=red,anchorcolor=red,citecolor=green]{hyperref}
\usepackage{setspace}
\usepackage{booktabs}
\usepackage{float}

\begin{document}

\thispagestyle{empty}

\begin{center}

\title{Rate of the phase transition for a charged anti-de Sitter black hole}

\author{Zhen-Ming Xu\footnote{E-mail: zmxu@nwu.edu.cn}, Bin Wu\footnote{E-mail: binwu@nwu.edu.cn}, and Wen-Li Yang\footnote{E-mail: wlyang@nwu.edu.cn}
        \vspace{6pt}\\}

\affiliation{$^{1}$School of Physics, Northwest University, Xi'an 710127, China\\
$^{2}$Institute of Modern Physics, Northwest University, Xi'an 710127, China\\
$^{3}$Shaanxi Key Laboratory for Theoretical Physics Frontiers, Xi'an 710127, China\\
$^{4}$Peng Huanwu Center for Fundamental Theory, Xi'an 710127, China}

\begin{abstract}
Phase transition is a core content of black hole thermodynamics. This study adopted the Kramer's escape rate method for describing the Brownian motion of particles in an external field to investigate the intensity of the phase transition between small and large black hole states. Some existing studies mostly focused on the formal analysis of the thermodynamic phase transition of black holes, but they neglected the detailed description of the phase transition process. Our results show that the phase transition between small and large black holes for charged anti-de Sitter (AdS) black holes presents serious asymmetric features, and the overall process is dominated by the transition from a small black hole to a large black hole. This study filled a research gap of a stochastic process analysis on the issue of the first-order phase transition rate in the AdS black hole.
\end{abstract}

\maketitle
\end{center}



\section{Introduction}
At present, black hole thermodynamics is considered one of the platforms for testing the quantum properties of black holes. In particular, the thermal entropy of a black hole, which is proportional to its area of event horizon~\cite{Bekenstein1973}, has attracted extensive discussion and interest. After decades of development, the thermal physics of black holes began to have a relatively complete theoretical framework. Although many issues still need to be further explained, black hole thermodynamics can be regarded as an effective window to explore quantum gravity. Of particular interest is the phase transition of black holes in the anti-de Sitter (AdS) spacetime, which closely links statistical mechanics, quantum mechanics, and general relativity. The most representative is the Hawking–Page phase transition between thermal radiation and the large AdS black hole~\cite{Hawking1983}, which is elaborated as the confinement/deconfinement phase transition in a gauge field~\cite{Witten1998}. The introduction of an extended phase space has injected extremely high vitality into the study of black hole thermodynamics~\cite{Kastor2009,Dolan2011,Chamblin1999,Bhattacharya2017a,Hendi2016,Cai2013,Spallucci2013,Zangeneh2018,Kubiznak2012,Kubiznak2017}. Through the analogy analysis with van der Waals fluid, people believe that the (theoretical) black hole has a certain microstructure~\cite{Wei2015,Wei2019,Wei2020,Miao2018,Xu2020,Ghosh2020,Bhattacharya2017b}. Meanwhile, some recent works~\cite{Zhang2015,Visser2022,Cong2021,Cong2022,Gao2022a,Zhao2022,Gao2022b,Kong2022,Bai2022} are devoted to using the anti-de Sitter/conformal field theory correspondence to explore the physical explanation of black hole thermodynamics in the field theory and make the relevant concepts of black hole thermodynamics reasonably and reliably.

Existing studies tend to focus on the type and criticality analysis of the thermodynamic phase transition of black holes, but they neglect the detailed description of the phase transition process. Recently, the free energy landscape has been proposed to explore some related evolution processes of the black hole phase transition under the background of the non-equilibrium statistical physics~\cite{Li2020a,Li2020b,Li2021,Wei2021,Cai2021}. By calculating the mean first passage time, these studies have preliminarily investigated some kinetics of the phase transition. In addition, the point of view of Landau free energy was used for the same analysis~\cite{Xu2021a}. Because some thermodynamic processes in a thermodynamic system are driven by stochastic fluctuations, the use of the analysis method of relevant stochastic processes in non-equilibrium statistical physics to obtain important information about the occurrence of thermodynamic processes is also reasonable.

In this study, we investigated the black hole phase transition rate. The black hole phase transition we mentioned here refers to the large and small black hole phase transitions in the AdS black hole, which is similar to the gas–liquid phase transition in van der Waals fluid~\cite{Kubiznak2012}. The phase transition of large and small black holes is dynamic. That is, the transition from a large black hole to a small black hole will be accompanied by the one from a small black hole to a large black hole, and vice versa. Naturally, two key issues need to be clarified:
\begin{itemize}
  \item Which of the two processes dominates in the phase transition of the AdS black hole?
  \item Under what circumstances will the two processes achieve dynamic balance?
\end{itemize}

In this letter, we will try to explain the two issues to deepen our understanding of the process of the large and small black hole phase transitions in charged AdS black holes, that is, the first-order phase transition. In equilibrium statistical physics, based on the behavior of Gibbs free energy, we can determine the state that the system tends to be in with the change in the temperature in the thermodynamics process, and the intersection of a swallow tail structure in Gibbs free energy is the turning point of which preferred state the system is in~\cite{Huang1987,Johnston2014}. However, these are all result descriptions, and the process during the period needs further in-depth analysis. This case is the main motivation for this study.

Next, we will take the charged AdS black hole, which possesses the most typical thermodynamic phase transition of a black hole, as an example to discuss. Using the Kramer's escape rate method~\cite{Risken1988,Zwanzig2001} of describing the Brownian motion of particles in an external field and thermal potential we constructed in~\cite{Xu2021b}, we examine some process characteristics in the first-order phase transition of the black hole. The results indicate that the rate of the first-order phase transition of the black hole shows a trend of increasing first and then decreasing, and the rate of transition from a small black hole to a large black hole is much greater than that from a large black hole to a small black hole within a wide range of ensemble temperatures. Two processes (the transition from a large black hole to a small black hole and the one from a small black hole to a large black hole) present serious asymmetric features, and the overall process is dominated by the transition from a small black hole to a large black hole.

\section{Charged AdS black hole}
We briefly review the thermodynamic behavior of the charged AdS black hole~\cite{Kubiznak2012}. The mass of the black hole in terms of the radius of the event horizon $r_h$ is
\begin{eqnarray}\label{enthalpy}
M=\frac{r_h}{2}+\frac{4\pi P r_h^3}{3}+\frac{Q^2}{2r_h^2},
\end{eqnarray}
where $Q$ is the total charge of the black hole, $P$ is the thermodynamic pressure defined by $P=3/(8\pi l^2)$, and $l$ is the AdS radius. The temperature of the charged AdS black hole is
\begin{eqnarray}\label{temperature}
T_h=\frac{1}{4\pi r_h}\left(1+8\pi P r_h^2-\frac{Q^2}{r_h^2}\right),
\end{eqnarray}
and the entropy conjugated with the temperature is
\begin{eqnarray}\label{entropy}
S=\pi r_h^2.
\end{eqnarray}

Gibbs free energy plays a key role in the phase transition of a thermodynamic system. For the charged AdS black hole, it is
\begin{eqnarray}
G\equiv M-T_h S=\frac{1}{4}\left(r_h-\frac{8\pi}{3}P r_h^3+\frac{3Q^2}{r_h}\right).
\end{eqnarray}

Several studies have shown that the thermodynamic behavior of a charged AdS black hole is similar to that of van der Waals fluid. For van der Waals fluid, it undergoes the gas–liquid phase transition, and for the charged AdS black hole, it is called the large–small black hole phase transition. The critical values for the black hole are~\cite{Kubiznak2012}
\begin{eqnarray}\label{criticals}
r_c=\sqrt{6}Q, \quad T_c=\frac{\sqrt{6}}{18\pi Q}, \quad P_c=\frac{1}{96\pi Q^2}, \quad G_c=\sqrt{6}Q/3.
\end{eqnarray}

For the convenience of discussion, we now introduce dimensionless thermodynamic quantities, which are respectively defined as follows:
\begin{eqnarray}\label{reparameters}
t_h:=\frac{T_h}{T_c}, \quad p:=\frac{P}{P_c}, \quad x:=\frac{r_h}{r_c}, \quad g:=\frac{G}{G_c}.
\end{eqnarray}

In our previous work~\cite{Xu2021b}, we consider a canonical ensemble composed of a large number of states (on-shell black hole states and off-shell other unknown states) and construct the thermal potential:
\begin{eqnarray}\label{tpotentiald}
f(x)=\int (T_h-T)\text{d}S.
\end{eqnarray}
Here, when the ensemble temperature $T$ is equal to the Hawking temperature $T_h$, the ensemble is made up of an on-shell black hole state and is in equilibrium. Moreover, when $T\neq T_h$, all possible states in the canonical ensemble deviate from the on-shell black hole state. In usual thermodynamics, the degree of thermal motion is measured by the product of the temperature and entropy. Hence, the above thermal potential roughly reflects the degree of deviation. Moreover, we reconstructed the phase transition behavior of black holes using the geometric characteristics of such an external potential. The ensemble temperature $T$ can take any positive value in any way, whereas the temperature $T_h$ of the on-shell black hole states can also take any positive value based on Eq.~(\ref{temperature}). Therefore, $T=T_h$ is just one of the ways to derive the value of the ensemble temperature $T$. Driven by thermal fluctuations, black hole states move in such a thermal potential. The dynamic behavior of black holes in the thermal potential can reflect some thermodynamic phase transition characteristics of black holes, which is consistent with the analysis given by Gibbs free energy.

\begin{figure}[htb]
\begin{center}
\subfigure[swallow tail]{\includegraphics[width=50 mm]{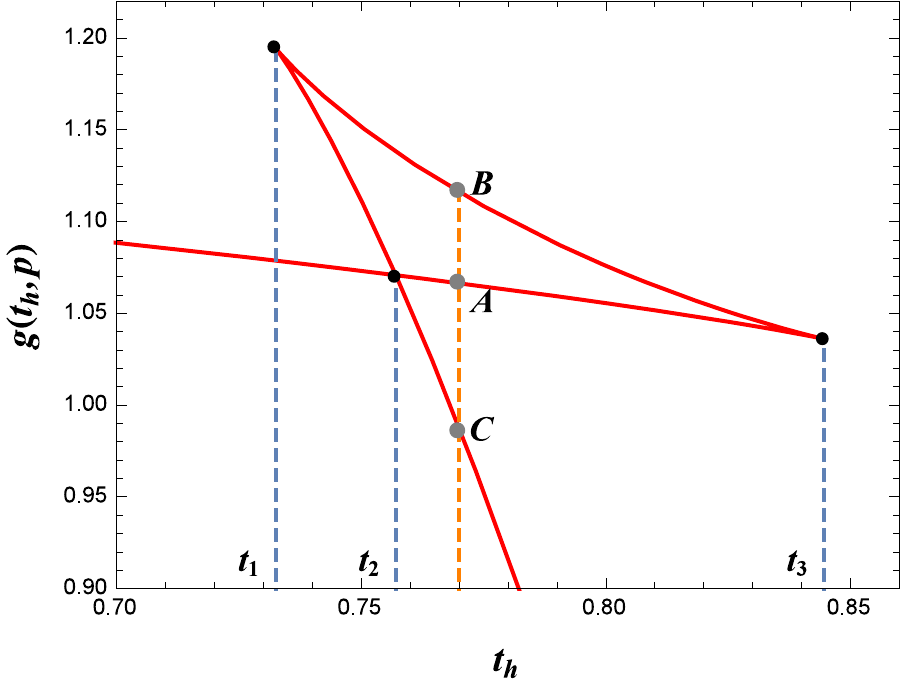}}
\qquad
\subfigure[$\omega$-well potential ]{\includegraphics[width=50 mm]{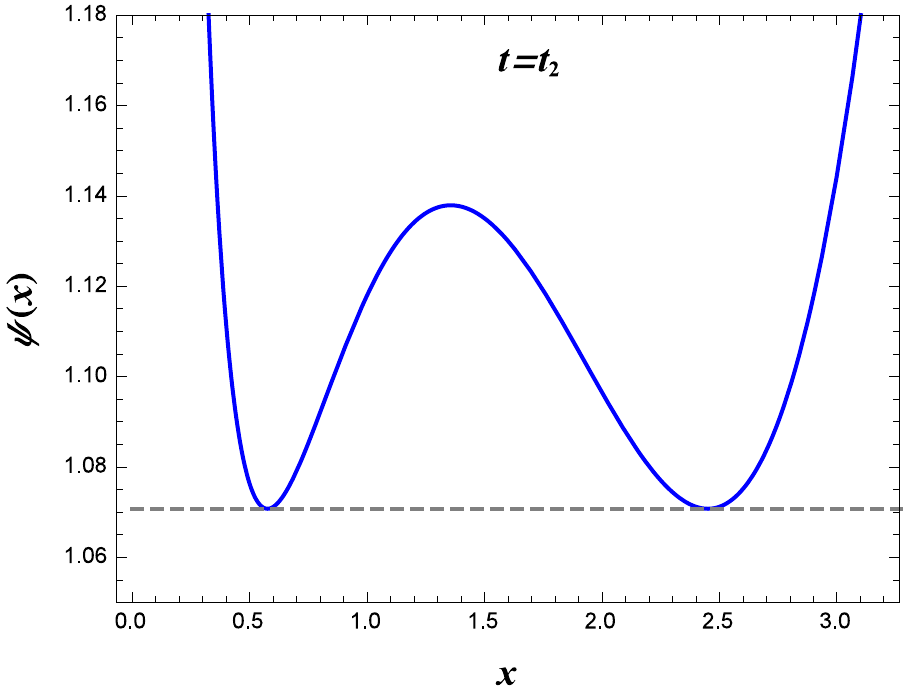}}
\qquad
\subfigure[double-well potential]{\includegraphics[width=50 mm]{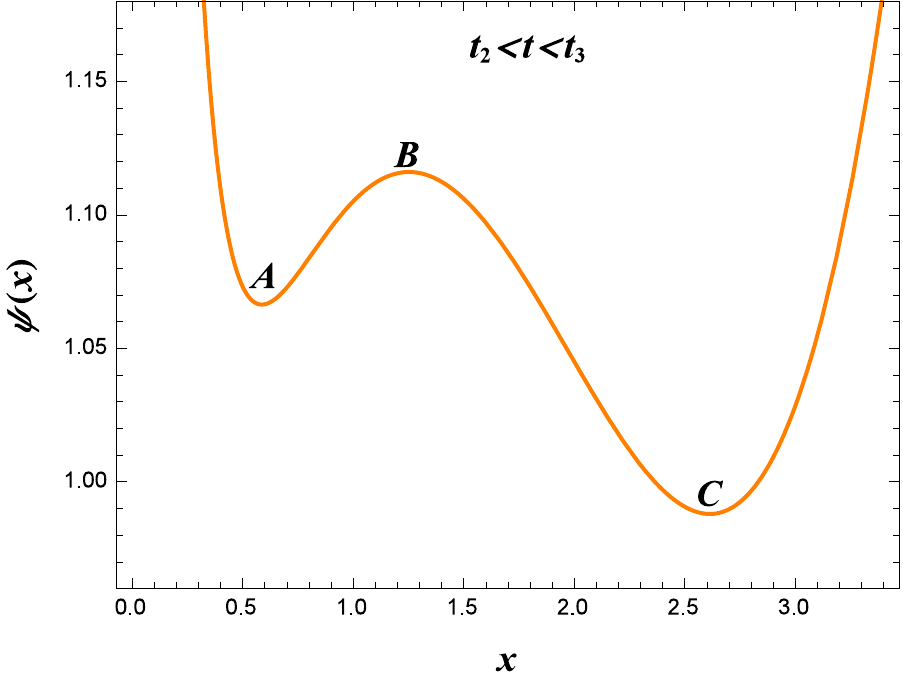}}
\end{center}
\caption{Behaviors of the dimensionless Gibbs free energy $g(t_h,p)$ and dimensionless thermal potential $\psi(x)$ for the charged AdS black hole when the thermodynamic pressure $p=\text{constant}<1$.}
\label{fig1}
\end{figure}

By introducing the external potential, we can learn more about the properties of the black hole phase transition from the geometric characteristics of the thermal potential.
Substituting Eqs.~(\ref{temperature}),~(\ref{entropy}),~(\ref{criticals}) and~(\ref{reparameters}) into Eq.~(\ref{tpotentiald}) and completing the integration, we can obtain the expression of the thermal potential for the charged AdS black hole:
\begin{eqnarray}\label{tpotential}
f(x)=\frac{\sqrt{6}Q}{3}\psi(x)=\frac{\sqrt{6}Q}{3}\left(\frac{1}{4x}+\frac{3x}{2}+\frac{px^3}{4}-tx^2\right),
\end{eqnarray}
where we replace the ensemble temperature $T$ with its dimensionless version $t$ via $t:=T/T_c$. In FIG.~\ref{fig1}, we show the behaviors of a dimensionless Gibbs free energy $g(t_h,p)$ and dimensionless thermal potential $\psi(x)$. Here, we notice three key values of the temperature~\cite{Xu2021a}:
\begin{eqnarray}\label{t13}
   t_1=\frac{3-p+3\sqrt{1-p}}{2p\left(\frac{\sqrt{1-p}+1}{p}\right)^{3/2}}, \quad t_2=\sqrt{\frac{p(3-\sqrt{p})}{2}}, \quad
   t_3=\frac{ \left(2-\sqrt{1-p}\right)\sqrt{1+\sqrt{1-p}}}{2},
\end{eqnarray}
where $t_1$ and $t_3$ are the local minimum temperature and the local maximum temperature of the swallowtail tip, and $t_2$ is the temperature of the swallowtail intersection. A thermodynamic system is always inclined to be in the state with the lowest Gibbs free energy. When the temperature is between $t_1$ and $t_3$, the system is in three possible states, marked as $A$, $B$, and $C$. In terms of equivalence, the thermodynamic system at this time is in the thermal potential shown in diagrams (b) and (c) in FIG.~\ref{fig1}. When the ensemble temperature $t=t_2$, the two global minima in the thermal potential are equal, labeled as $\omega$-well. When the ensemble temperature $t\in (t_1, t_3)$ (excluding $t_2$), the thermal potential presents the characteristics of a general double-well potential.

Based on the black hole molecular hypothesis~\cite{Wei2015}, the large–small black hole phase transition is the rearrangement of black hole molecules due to thermal fluctuations in the thermal potential. Diagram (c) in FIG.~\ref{fig1} might describe a molecular rearrangement. We assume that the ensemble temperature (multiplied by the Boltzmann factor) is much lower than the barrier height. Molecules will spend a lot of time near the potential minimum (point $A$), and only rarely will Brownian motion take them to the top of the barrier (point $B$). Once the molecule reaches the top of the barrier, it is likely to fall equally to either side of the barrier. If it moves to the right-hand side, it will rapidly fall to the other minimum (point $C$), stay there for a while, and then perhaps cross back to the original minimum (point $A$).

Here, the motion is purely diffusive, governed by a Smoluchowski equation, and the barrier is high (or the temperature is low). Hence, it is quite easy to find a crossing rate, which is the Kramer's rate~\cite{Risken1988,Zwanzig2001}:
\begin{eqnarray}\label{rate}
r_k=\dfrac{\sqrt{|f^{''}(x_{\text{min}})f^{''}(x_{\text{max}})|}}{2\pi}e^{-\dfrac{f(x_{\text{max}})-f(x_{\text{min}})}{D}},
\end{eqnarray}
where $D$ is the constant diffusion coefficient, $x_{\text{max,min}}$ are the locations of the extreme points of the thermal potential, and the two prime represents the second derivative of the potential function $f(x)$ with respect to the parameter $x$.

\begin{figure}[htb]
\begin{center}
\includegraphics[width=65 mm]{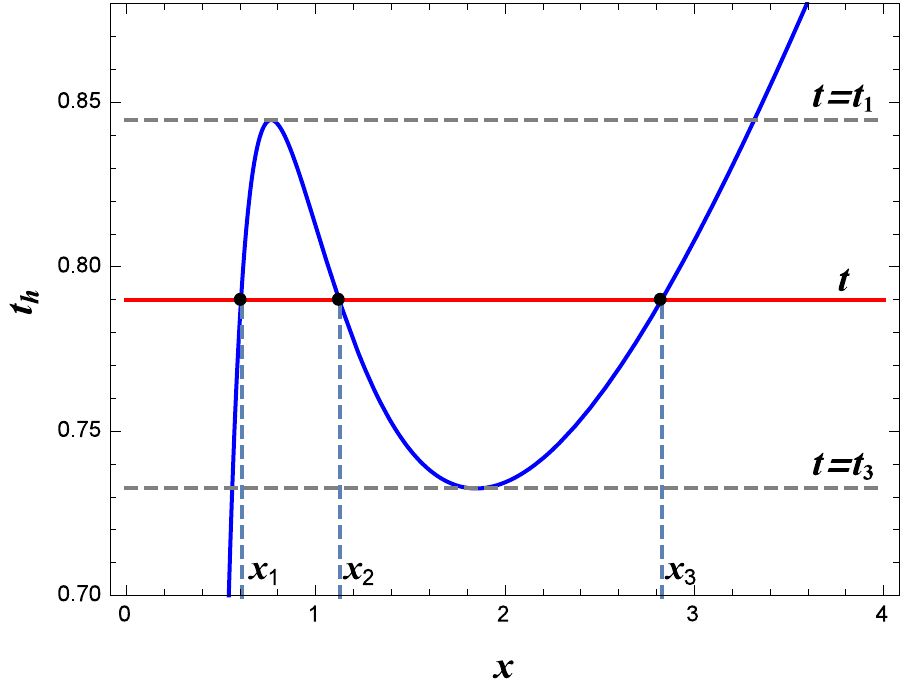}
\end{center}
\caption{Graphic representation of three real roots ($x_1$, $x_2$, and $x_3$) of an equation $t_h=t$ at the pressure $p=0.5$ for the charged AdS black hole.}
\label{fig3}
\end{figure}

For the charged AdS black hole, the locations of the extreme points of the thermal potential are determined by the equation $3px^4-8tx^3+6x^2-1=0$. When $0<p<1$ and $t\in (t_1, t_3)$, this equation has three real positive roots. From small to large, we mark them as $x_1$, $x_2$, and $x_3$. According to Eq.~(\ref{tpotentiald}), these positions, in essence, are the solutions of $T_h=T$ or $t_h=t$. With the dimensionless quantities~(\ref{reparameters}), we present a plot to illustrate this point in FIG.~\ref{fig3}, in which we can easily see the separations between the three positions.

Hence, the rate of crossing from $A$ to $C$ is labeled as $r_{k1}$, and the rate of crossing from $C$ to $A$ is labeled as $r_{k2}$, which can be read as
\begin{eqnarray}\label{rate12}
r_{k1}&=\dfrac{\sqrt{|f^{''}(x_1)f^{''}(x_2)|}}{2\pi}e^{-\dfrac{f(x_2)-f(x_1)}{D}},\\
r_{k2}&=\dfrac{\sqrt{|f^{''}(x_3)f^{''}(x_2)|}}{2\pi}e^{-\dfrac{f(x_2)-f(x_3)}{D}}.
\end{eqnarray}

\begin{figure}[htb]
\begin{center}
\subfigure[~transition rate]{\includegraphics[width=60 mm]{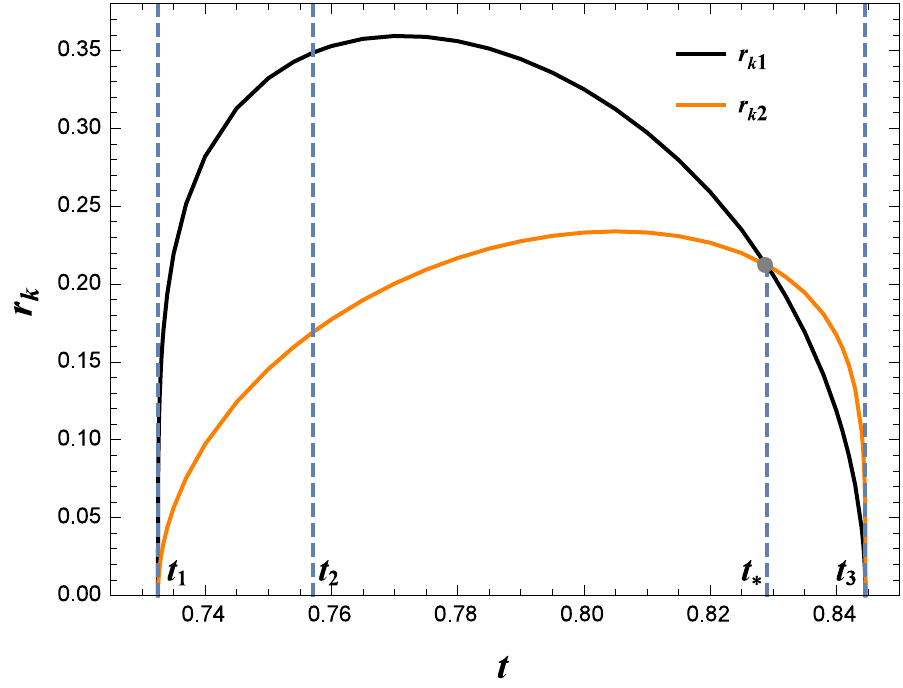}}
\qquad
\subfigure[~net transition rate]{\includegraphics[width=60 mm]{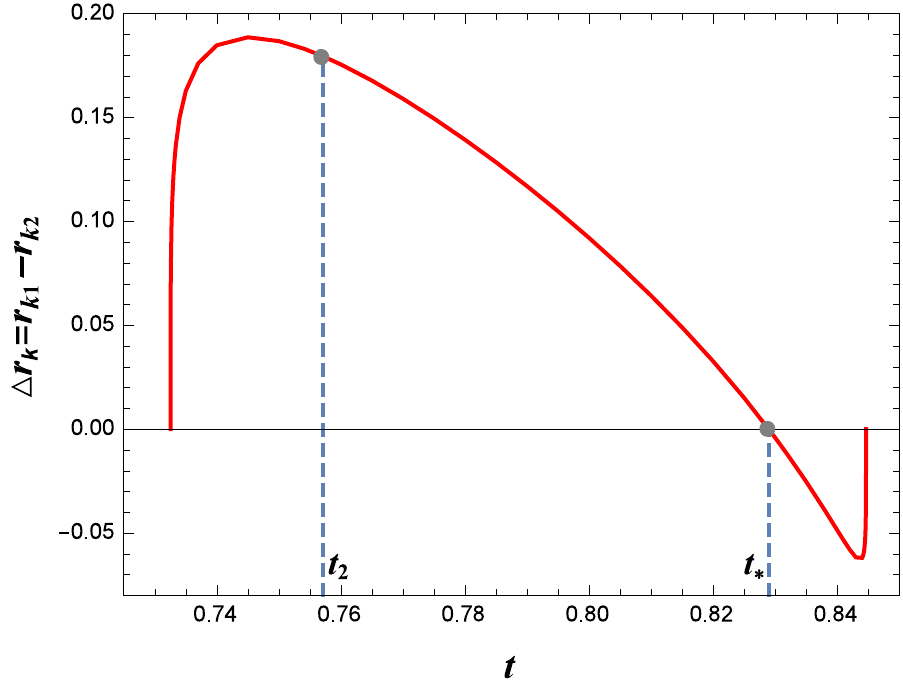}}
\end{center}
\caption{Transition rate $r_k$ with respect to the ensemble temperature $t$ at the pressure $p=0.5$ for the charged AdS black hole. Here without losing generality, we set the charge $Q=10/\sqrt{6}$ and constant diffusion coefficient $D=10$.}
\label{fig2}
\end{figure}

We plot the transition rate between different states of the charged AdS black hole in FIG.~\ref{fig2}, that is, the transition rate $r_{k1}$ from the small black hole state to the large black hole state and the transition rate $r_{k2}$ from the large black hole state to the small black hole state. Here, when using Eq.~(\ref{rate}) to analyze the transition rate, we assume that the ensemble temperature is much lower than the barrier height, $t T_c \ll \Delta f$, where $\Delta f$ is the barrier height, i.e.,
\begin{eqnarray}
\frac{\sqrt{6}}{18\pi Q}t \quad\ll\quad \frac{\sqrt{6}Q}{3}\Delta\psi
\end{eqnarray}
Therefore, we can always find the appropriate value of $Q$ to make the above condition meet.

From FIG.~\ref{fig2}, we can read some microscopic information about the phase transition of the charged AdS black hole between the small black hole state and the large black hole state.

At $t_1$ and $t_3$, the transition rate is zero, which means that no phase transition occurs. As the ensemble temperature $t$ increases from $t_1$ to $t_3$, the two rates show a trend of increasing first and then decreasing.

At $t_2$, according to the diagram (b) in FIG.~\ref{fig1}, the two global minima in the thermal potential are equal, but the transition rates $r_{k1}$ and $r_{k2}$ are not equal. The rate $r_{k1}$ is much larger than the rate $r_{k2}$, indicating that the transition process from the small black hole to the large black hole is far from that of the transition from the large black hole to the small black hole. Between $t_2$ and $t_*$, the two rates will reach a maximum value: $r_{k1}$ will reach first, and $r_{k2}$ will reach later. Before $t_*$, the net rate of the transition ($\Delta r_{k}=r_{k1}-r_{k2}$) between the two phases shows a trend of increasing first and then decreasing, and the rate of the transition $r_{k1}$ from the small black hole to the large black hole is dominant. At the same time, the net rate reaches the maximum value between $t_1$ and $t_2$.

At $t_*$, the transition rates $r_{k1}$ and $r_{k2}$ are equal, and the net rate is zero, which means that the transition between the two phases reaches dynamic equilibrium. Once the ensemble temperature $t$ exceeds $t_*$, there will be a reversal. That is, at this time, the process of the transition from the large black hole to the small black hole will dominate, and it also shows a trend of first increasing and then decreasing. When the ensemble temperature reaches $t_3$, both rates are zero, indicating that there is no phase transition.

In short, the phase transition between small and large black holes presents very asymmetric features, and the overall process is dominated by the transition from a small black hole to a large black hole.

\section{Summary}
In this study, we used the geometric behavior of a black hole in the thermal potential and adopted the Kramer's escape rate method of describing the Brownian motion of particles in an external field to study the dynamic process of the first-order phase transition for a charged AdS black hole. In the previous process of analyzing the phase transition behavior of black holes using Gibbs free energy, we cannot extract information on the transition intensity from one state to another. By means of a stochastic process, at present, we have obtained the rate behavior of the phase transition between small and large black holes for the charged AdS black hole.

On the whole, the first-order phase transition rate of the AdS black hole shows a trend of increasing first and then decreasing, and the transition rate from a small black hole to a large black hole is much greater than that from a large black hole to a small black hole within a wide range of ensemble temperatures. In the analysis of Gibbs free energy, there are three key temperatures $t_1$, $t_2$, and $t_3$ (see diagram (a) in FIG.~\ref{fig1}). Only when the ensemble temperature is between $t_1$ and $t_3$ can the phase transition occur. When the ensemble temperature $t=t_2$, it is at the swallow tail intersection of Gibbs free energy, which corresponds to the case where the two global minima in the thermal potential are equal (see diagram (b) in FIG.~\ref{fig1}). Intuitively, as the two minima are equal, the molecules at the two positions should simultaneously reach the top of the barrier, and the phase transition rate will be equal. In fact, at this temperature $(t=t_2)$, the transition rate from a small black hole to a large black hole is much faster than that of a large black hole to a small black hole (see FIG.~\ref{fig2}). This is the important information about the phase transition of black holes that we cannot get in the analysis of Gibbs free energy. Only when the ensemble temperature $t=t_*$, where $t_*\in (t_2, t_3)$, can the two rates be equal, indicating that the phase transition reaches a dynamic equilibrium. These results mean that the phase transition between small and large black holes for a charged AdS black hole presents serious asymmetric features, and the overall process is dominated by the transition from a small black hole to a large black hole.

In addition, the thermal potential cannot be arbitrary, and it should meet some requirements. The extreme value of the thermal potential corresponds to the equilibrium state, i.e., the state with the ensemble temperature $T$ equating to the Hawking temperature $T_h$. The concavity and convexity of the thermal potential can be related to the stability of the thermodynamic system. For the thermal potential under the two conditions, the expression may not be the same as that constructed in this study. Meanwhile, for the analysis of the phase transition rate, the qualitative behaviors are consistent, but there are differences in the numerical values.

Because the thermodynamic system is driven by fluctuations, we present the characteristics of the first-order phase transition rate of black holes using some analytical methods of the stochastic process. However, some problems still need to be further explained. What is the asymmetry mechanism in the phase transition rate between small and large black holes for a charged AdS black hole? Why is the phase transition rate between small and large black holes at the swallow tail intersection of Gibbs free energy not equal? These are very meaningful research contents. In addition, the current method can be extended to other types of phase transitions, especially the behavior near the triple points among black hole thermodynamic phase transitions, which will further understand the microscopic mechanism of black hole phase transitions.

\section*{Acknowledgments}
This research is supported by National Natural Science Foundation of China (Grant No. 12105222, No. 12275216, and No. 12247103), and also supported by The Double First-class University Construction Project of Northwest University. The author would like to thank the anonymous referee for the helpful comments that improve this work greatly.

\end{document}